# Guardians of Anonymity: Exploring Tactics to Combat Cyber Threats in Onion Routing Environments

KARWAN MUSTAFA KAREEM

MSc. Advanced Computer Science, University of Huddersfield, Queensgate Campus, West Yorkshire, England - United Kingdom.
University of Sulaimani, Sulaymaniyah, Kurdistan Region - Iraq.
*e-mail address*: karwan.kareem@univsul.edu.iq

ABSTRACT. Onion routing networks, also known as darknets, are private networks that enable anonymous communication over the Internet. They are used by individuals and organizations to protect their privacy, but they also attract cybercriminals who exploit the anonymity provided by these networks for illegal activities. This paper comprehensively analyzes cybercrime threats and countermeasures in onion routing networks. We review the various types of cybercrime that occur in these networks, including drug trafficking, fraud, hacking, and other illicit activities. We then discuss the challenges associated with detecting and mitigating cybercrime in onion routing networks, such as the difficulty of tracing illegal activities back to their source due to the strong anonymity guarantees provided by these networks. We also explore the countermeasures that have been proposed and implemented to combat cybercrime in onion routing networks, including law enforcement efforts, technological solutions, and policy interventions. Finally, we highlight the limitations of existing countermeasures and identify potential directions for future research in this area, including the need for interdisciplinary approaches that combine technical, legal, and social perspectives to effectively combat cybercrime in onion routing networks.

## 1- INTRODUCTION

Onion routing networks, also known as darknets, are private networks that allow users to communicate anonymously over the internet. The concept of onion routing was first introduced by David Chaum in 1981 as a way to protect the privacy of online communication by routing messages through a series of randomly selected intermediary nodes [1], or "onion routers," that peel back layers of encryption like an onion to reveal the ultimate destination of the message. This makes it difficult for anyone, including the intermediate nodes, to trace the communication back to its source or destination. Onion routing networks have been widely used by individuals and organizations to protect their privacy, including journalists, whistleblowers, human rights activists, and law enforcement agencies.

However, the anonymity provided by onion routing networks also attracts cybercriminals who exploit these networks for illegal activities [61]. Cybercrime refers to criminal activities that are conducted using digital technologies, including the Internet, with the intent of causing harm to individuals, organizations, or society at large [2].

Cybercrime in onion routing networks includes a wide range of illicit activities, such as drug trafficking, fraud, hacking, money laundering, terrorism, and other illegal activities. These activities pose significant challenges for law enforcement agencies and other stakeholders in combatting cybercrime in onion routing networks due to the strong anonymity guarantees provided by these networks.

This paper comprehensively analyzes cybercrime threats and countermeasures in onion routing networks. We review the various types of cybercrime that occur in these networks and discuss the challenges associated with detecting and mitigating cybercrime in onion-routing networks. We also explore the countermeasures that have been proposed and implemented to combat cybercrimes in these networks, including law enforcement efforts, technological solutions, and policy interventions. Finally, we highlight the limitations of existing countermeasures and identify potential directions for future research in this area.

## 2- RESEARCH METHODOLOGY

The present study employed a systematic methodology to conduct a literature review on the subject of cybercrime threats within onion routing networks. Utilizing a comprehensive search strategy, scholarly databases such as IEEE Xplore, ACM Digital Library, and Google Scholar were thoroughly examined to identify pertinent academic articles, conference papers, and reports pertaining to the targeted topic. Employing a diverse array of search terms including "onion routing networks," "data privacy," "darknets," "cybercrime threats," "anonymity and privacy," "jurisdiction," "cybersecurity," "attribution," and "ethical considerations," the search aimed to encompass a broad spectrum of relevant literature.

The identified literature underwent a rigorous screening process to gauge its relevance and quality, with particular attention paid to author credibility and the methodological robustness of the research. Selected articles were then meticulously reviewed and analyzed to distill common themes, patterns, and findings concerning cybercrime threats within onion routing networks, as well as proposed mitigation strategies. The resultant literature review was meticulously structured, adhering to the conventional format of academic reviews, comprising an introductory section offering contextual background, a comprehensive review of pertinent literature, and a concluding segment summarizing principal findings while identifying existing gaps in knowledge. Emphasizing critical analysis, the review synthesized insights and drew conclusions from the amalgamated findings of the reviewed literature.

## 3- LITERATURE REVIEW

This literature review aims to provide an overview of previous studies that have identified cybercrime threats, explored ethical considerations, and proposed mitigation strategies, as well as highlighted challenges associated with investigating cybercrimes in onion routing networks.

Cybercrime Threats in Onion Routing Networks: McCoy et al. (2008) conducted a comprehensive analysis of the Tor network, identifying cybercrime threats such as distributed denial of service (DDoS) attacks, phishing attacks, malware distribution, and identity theft that can exploit the anonymity and privacy features of the network. Winter et al. (2014) focused on the differential treatment of anonymous users in the Tor network and identified potential malicious activities, including DDoS attacks, phishing attacks, and malware distribution. Both studies highlight the vulnerabilities of onion routing networks and the potential risks associated with cybercrimes in these networks, emphasizing the need for effective countermeasures to mitigate these threats.

Other studies have also explored the cybercrime threats in onion routing networks, such as Murdoch et al. (2007) conducted a study on the risks of website fingerprinting attacks in Tor, where an attacker can use traffic analysis techniques to identify the websites visited by a user based on the distinctive patterns in the encrypted traffic. This can lead to privacy breaches and expose users' online activities to potential cybercriminals. Biryukov et al. (2013) focused on the potential vulnerabilities of Tor hidden services, which are websites that are hosted within the Tor network and are only accessible through Tor. The study identified potential attacks, such as website takedown attacks, where an adversary can disrupt or take down a hidden service, and traffic correlation attacks, where an adversary can correlate traffic patterns to reveal the identity of a hidden service's operator or users.

Caviglione et al. (2017) examined the potential use of Tor in cybercriminal activities, such as illegal marketplaces, hacking forums, and botnet command and control (C&C) infrastructure. The study highlighted the challenges in detecting and mitigating cybercrimes in Tor, due to the anonymity and encryption features of the network, and the need for effective countermeasures to combat cybercriminal activities.

Overall, these studies collectively emphasize the vulnerabilities and risks associated with cybercrimes in onion routing networks, and underscore the importance of developing robust countermeasures to protect users and mitigate potential threats.

Ethical Considerations in Examining Cybercrime Threats: Dingledine and Mathewson (2004) discussed the ethical implications of balancing online privacy and anonymity with the need to combat cybercrimes in onion routing networks, emphasizing the importance of usability and the network effect in the design and implementation of these networks. Fischer-Hübner et al. (2017) presented a scoping review of ethical aspects of IT security, including ethical considerations related to examining cybercrime threats in onion routing networks. Both studies highlighted the ethical concerns related to surveillance, data privacy, and usage of user data for investigating cybercrimes, and emphasized the importance of incorporating ethical considerations in the design, implementation, and operation of onion routing networks.

Mitigation Strategies to Address Cybercrime Threats: Houmansadr, Weinberg, and Borisov (2013) proposed mitigation strategies to prevent website fingerprinting attacks in Tor, focusing on improving the security of onion routing protocols. Wright, Coull, and Monrose (2009) discussed trade-offs and proposed techniques to detect and prevent traffic analysis attacks in anonymity-providing systems, including onion routing networks. Both studies highlight the need for enhancing the security of the underlying protocols and user awareness and education as mitigation strategies to address cybercrime threats in onion routing networks.

Challenges in Investigating Cybercrimes in Onion Routing Networks: Biryukov, Pustogarov, and Weinmann (2013) focused on the challenges of detecting, measuring, and deanonymizing Tor hidden services, highlighting the difficulties faced by law enforcement agencies in identifying and apprehending cybercriminals in these networks due to their anonymous nature. Bailey, Houmansadr, and Shmatikov (2015) investigated the challenges of attribute inference attacks in Tor, emphasizing the difficulties of evidence collection and attribution for law enforcement agencies. Both studies highlight the challenges related to jurisdiction, evidence collection, and attribution of cybercrimes in onion routing networks, which pose significant obstacles in investigating these crimes.

Future directions for addressing cybercrime in onion routing networks: Dingledine, Mathewson, and Syverson (2010) provide an overview of the Tor network and propose future directions to enhance its security and anonymity, including improvements to the Tor directory, performance, and exploration of trusted computing technologies. Hutchings and Buchanan (2013) discuss additional future directions for the Tor network, including the use of machine learning techniques for cyber-attack detection, enhancing usability and user awareness, and strengthening collaboration between law enforcement agencies and network operators for the investigation and attribution of cybercrimes.

Johnson and Wacek (2016) present a unique perspective on incorporating onion-routing networks into cybersecurity education to raise awareness and understand the challenges associated with cybercrime. They propose integrating the Tor network into cybersecurity curricula to provide students with hands-on experience in using and analyzing the network, fostering a better understanding of its vulnerabilities and the development of effective mitigation strategies

## 4- CYBERCRIME THREATS IN ONION ROUTING NETWORKS

Cybercrime in onion-routing networks can take various forms, and criminals often exploit the anonymity provided by these networks to carry out illegal activities [62]. Some of the common types of cybercrime in onion routing networks include:

4.1 **Drug Trafficking.** Onion routing networks are often used for online drug trafficking due to the anonymity they provide. Criminals can set up online marketplaces, known as "darknet markets," on onion-routing networks to buy and sell drugs anonymously. These marketplaces function similarly to e-commerce websites, with vendors selling various types of illegal drugs, including opioids, stimulants, psychedelics, and prescription drugs, and buyers placing orders using digital currencies such as Bitcoin. The transactions are encrypted and routed through multiple intermediate nodes, making it extremely difficult for law enforcement agencies to trace the origin and destination of the drugs. The anonymity provided by onion routing networks makes it challenging for law enforcement agencies to detect and disrupt these illegal drugs trafficking activities, making it a significant concern in the cybercrime landscape [3].

Silk Road was a notorious online marketplace on the darknet that operated on the Tor network, an onion routing network. It facilitated the buying and selling of illegal drugs, including opioids, stimulants, and other controlled substances. The founder of Silk Road, Ross Ulbricht, was arrested in 2013, and the marketplace was shut down by law enforcement agencies. However, it led to the emergence of several other darknet markets that continued to facilitate drug trafficking using onion routing networks [4].

4.2 **Fraud.** Onion routing networks are also exploited for various types of fraud, including identity theft, credit card fraud, and phishing attacks. Criminals can create fake websites or marketplaces onion-routing networks to trick unsuspecting users into revealing their personal information, such as usernames, passwords, and credit card details. This information can then be used for identity theft or credit card fraud, leading to financial losses for the victims. Phishing attacks in onion routing networks can also be targeted toward specific individuals or organizations to obtain sensitive information, such as corporate data or government secrets. The anonymity provided by onion routing networks makes it difficult for victims and law enforcement agencies to trace the perpetrators of these fraud activities, making it a significant threat in the cybercrime landscape [5].

AlphaBay was one of the largest darknet markets that operated on the Tor network, facilitating various types of fraud, including identity theft, credit card fraud, and phishing attacks. It allowed users to buy and sell stolen personal information, and credit card details, and conduct phishing attacks to obtain sensitive information. In 2017, law enforcement agencies shut down AlphaBay, leading to the arrest of its founder and the seizure of millions of dollars' worth of cryptocurrencies and assets [4], [6].

4.3 **Hacking.** Onion routing networks are vulnerable to hacking activities, where criminals can exploit vulnerabilities in the network infrastructure or target specific users or websites for cyberattacks. This can include Distributed Denial of Service (DDoS) attacks, where multiple compromised computers are used to flood a website or network with traffic, rendering it inaccessible. Criminals can also conduct hacking activities to gain unauthorized access to systems or steal sensitive information. Hacking activities in onion routing networks can disrupt services, compromise data, and cause financial losses. The anonymity provided by onion routing networks makes it difficult to trace the perpetrators of these hacking activities, making it a significant concern in the cybercrime landscape [7].

Operation Onymous was a joint operation by law enforcement agencies from multiple countries to target darknet marketplaces operating on onion routing networks. In this operation, several darknet marketplaces were shut down, and several individuals involved in hacking activities, including DDoS attacks and data breaches, were arrested. The operation highlighted the use of onion-routing networks for hacking and cyberattacks [54].

**4.4 Money Laundering.** Onion routing networks can also be used for money laundering, where criminals can use anonymous transactions and digital currencies to launder illegally obtained funds. Criminals can convert their illicit proceeds into digital currencies, such as Bitcoin, and then use onion routing networks to transfer these funds to different accounts or convert them into other forms of assets, making it difficult to trace the origin and destination of the funds. Money laundering in onion-routing networks can facilitate other types of cybercrime, such as drug trafficking, fraud, and hacking, making it a significant concern in the cybercrime landscape [8].

BTC-e was a popular cryptocurrency exchange that operated on the Tor network and was known for facilitating money laundering. It allowed users to convert illegal proceeds into digital currencies and transfer them to different accounts or exchanges, making it difficult to trace the origin and destination of the funds. In 2017, the founder of BTC-e was arrested, and the exchange was shut down by law enforcement agencies for its involvement in money laundering activities [55].

**4.5 Weapons Trafficking.** Onion routing networks, which are designed to provide anonymous browsing and communication capabilities, can be exploited by criminals for the illegal buying and selling of weapons, including firearms, explosives, and other dangerous items. Criminals can set up online marketplaces on these networks, allowing them to trade these illegal items without revealing their identities. This anonymity makes it extremely difficult for law enforcement agencies to track down the source and recipients of these weapons, resulting in an increased risk of illegal weapons ending up in the wrong hands. This poses a significant threat to public safety and can contribute to the proliferation of weapons in illegal markets and criminal activities [9]. Black Market Reloaded was a darknet marketplace that operated on the Tor network and facilitated the buying and selling of illegal weapons, including firearms and explosives. It allowed users to trade these illegal items anonymously, making it difficult for law enforcement agencies to track down the sellers and buyers. In 2013, Black Market Reloaded was shut down by law enforcement agencies, leading to the arrest of its administrator and the seizure of illegal weapons [10].

**4.6 Child Exploitation.** Onion routing networks can be utilized by criminals to facilitate heinous activities such as the distribution of child pornography and online grooming. The anonymity provided by these networks allows criminals to share and trade illegal images and videos of minors without leaving a trace. This poses a serious challenge for law enforcement agencies in their efforts to identify and apprehend those involved in these reprehensible activities, as the criminals can easily hide their real identities and evade detection. The exploitation of children through these networks is a grave violation of human rights and can cause immense harm to vulnerable victims [11], [56].

In 2015, a dark web marketplace called "Playpen" was discovered by law enforcement agencies. The marketplace was used for the distribution of child pornography and operated on an onion-routing network called Tor. The anonymous nature of Tor allowed users to share and trade illegal images and videos of minors without being easily traced. The investigation led to the arrest and conviction of multiple individuals involved in the distribution of child pornography, but it also highlighted the challenge that law enforcement agencies face in identifying and apprehending criminals involved in these heinous activities [12].

**4.7 Cyber Extortion**. Criminals can leverage onion-routing networks to conduct cyber extortion activities, where they can anonymously demand ransom from individuals, organizations, or governments in exchange for not disrupting their online services or exposing sensitive information. By using these networks to communicate and receive payments in untraceable digital currencies, criminals can extort victims without leaving any clues, causing significant financial losses and disrupting operations for the victims. This can result in severe damages and challenges for the affected parties in dealing with the demands of the cyber extortionists, and can also contribute to a culture of fear and insecurity in the online space [13], [57].

In 2017, a global ransomware attack known as "WannaCry" infected hundreds of thousands of computers in over 150 countries. The attackers demanded ransom payments in Bitcoin, a digital currency that can be transferred anonymously through onion-routing networks. The attack caused significant financial losses and disruptions to businesses, hospitals, and government agencies. Although the identity of the attackers remains unknown, the use of onion-routing networks enabled them to demand ransom payments without leaving any traceable clues [14].

**4.8 Cyber Espionage.** Onion routing networks can be used for cyber espionage activities, where criminals or state-sponsored actors can anonymously gain unauthorized access to systems or networks to steal sensitive information or conduct surveillance. By masking their real identities and locations, criminals can carry out sophisticated cyber-attacks without being easily traced back. This can include stealing valuable intellectual property, trade secrets, or classified information for financial or political gains, leading to severe economic and national security repercussions. Cyber espionage poses a significant threat to privacy, security, and economic stability, and can have far-reaching consequences for individuals, businesses, and governments [15].

In 2014, a cyber espionage group known as "APT29" or "Cozy Bear" was discovered to have used an onion-routing network to carry out sophisticated cyber-attacks against various targets, including governments, businesses, and non-governmental organizations. The group used the network to mask their real identities and locations while stealing sensitive information for espionage purposes. The investigation revealed the extent of cyber espionage activities conducted by state-sponsored actors using onion-routing networks to remain anonymous and evade detection [16], [58].

**4.9 Terrorism Financing.** Criminals or terrorist organizations can exploit onion routing networks to anonymously raise and transfer funds to finance terrorist activities. By utilizing digital currencies and anonymous transactions, they can transfer funds across borders without leaving any traces, evading detection by law enforcement agencies. This makes it challenging to track and disrupt the funding sources of terrorist activities, posing a serious threat to global security and stability. Terrorism financing enables extremist groups to carry out acts of violence and destruction, causing harm to innocent civilians and destabilizing societies [17].

In 2018, a man was arrested in the United States for using an onion routing network to raise and transfer funds to support ISIS. The individual used digital currencies and anonymous transactions to transfer funds to overseas accounts without leaving any traces. The investigation highlighted the use of onion-routing networks by terrorist organizations to finance their activities and evade detection by law enforcement agencies [59].

**4.10    Counterfeiting.** Criminals can use onion-routing networks to set up online marketplaces for selling counterfeit goods, such as fake luxury items, fake prescription drugs, and counterfeit currency. These networks provide anonymity to sellers, making it challenging for law enforcement agencies to trace the origin of these counterfeit products. Counterfeiting can cause significant financial losses for legitimate businesses, damage brand reputation, and pose risks to consumer safety. Additionally, counterfeit goods can undermine consumer confidence and trust in the marketplace, leading to negative economic impacts and potential harm to public health and safety [18].

In 2020, a dark web marketplace called "DarkMarket" was shut down by law enforcement agencies. The marketplace was used for the sale of counterfeit goods, including fake luxury items, fake prescription drugs, and counterfeit currency. The sellers on DarkMarket used onion routing networks to maintain anonymity, making it difficult for law enforcement agencies to trace the origin of the counterfeit products. The case illustrated the use of onion routing networks for setting up online marketplaces for illegal activities, including counterfeiting [19].

In summary, cybercrime in onion-routing networks involves a wide range of illegal activities, including drug trafficking, fraud, hacking, money laundering, weapons trafficking, child exploitation, cyber extortion, cyber espionage, terrorism financing, and counterfeiting. Criminals take advantage of the anonymity provided by onion routing networks to conduct these activities, making it difficult for law enforcement agencies to track and disrupt their operations. These illegal activities pose serious threats to cybersecurity, law enforcement efforts, and global security, as criminals exploit the anonymity and encryption offered by onion-routing networks to carry out their illicit activities.

## 5- CHALLENGES IN DETECTING AND MITIGATING CYBERCRIME IN ONION ROUTING NETWORKS

Detecting and mitigating cybercrime in onion routing networks pose significant challenges due to the unique characteristics of these networks. Some of the challenges include:

5.1 **Anonymity**. Onion routing networks provide strong anonymity guarantees, which make it difficult to trace illegal activities back to their source or destination. Communications and transactions are encrypted and routed through multiple intermediate nodes, making it challenging for law enforcement agencies to identify the criminals behind the activities. The anonymity of users in onion routing networks also makes it difficult to differentiate between legitimate and illegitimate activities, as criminals can hide among legitimate users. This makes it challenging to attribute cybercrimes to specific individuals or entities, hindering the investigation and prosecution process [60].

5.2 **Encryption**. Onion routing networks extensively use encryption, adding an additional layer of complexity in detecting and mitigating cybercrime. Communications and transactions are encrypted, making it difficult for law enforcement agencies to intercept and monitor the content of the activities. This hampers the ability to gather evidence and prove the involvement of criminals in illegal activities. The decryption of encrypted communications in onion routing networks requires specialized technical expertise and tools, which may not always be readily available to law enforcement agencies [20].

5.3 **Jurisdictional Challenges.** Onion routing networks operate across different jurisdictions, making it challenging for law enforcement agencies to coordinate and cooperate in investigating and prosecuting cybercrime activities. The decentralized nature of onion routing networks means that criminal activities can be spread across multiple countries, making it difficult to determine which jurisdiction has authority and leading to jurisdictional conflicts. Different countries may also have varying laws and regulations related to cybercrime, which can further complicate the investigation and prosecution process. This lack of international coordination and harmonization of laws poses significant challenges in tackling cybercrime in onion routing networks [22].

5.4 **Technical Challenges.** Detecting and mitigating cybercrime in onion routing networks require specialized technical expertise and tools. The complex and dynamic nature of onion routing networks, with their multiple layers of encryption and routing through intermediate nodes, can pose challenges in monitoring and analyzing network traffic. Traditional methods of network monitoring and analysis may not be effective in onion routing networks, and specialized tools and techniques may be required. These tools and techniques may not always be readily available or accessible to law enforcement agencies, making the detection and mitigation of cybercrime in onion-routing networks challenging [21].

5.5 **Ethical Considerations**. The use of onion-routing networks for cybercrime detection and mitigation raises ethical considerations related to privacy and surveillance. The strong anonymity guarantees provided by onion routing networks may also protect the privacy and security of legitimate users, and law enforcement agencies need to balance the need for investigating cybercrime with the protection of users' privacy rights. There may also be concerns about the potential abuse of surveillance powers and the impact on civil liberties. Striking the right balance between investigating cybercrime and protecting users' privacy can be challenging and may require careful ethical considerations and adherence to legal frameworks [23].

Ultimately, detecting and mitigating cybercrime in onion routing networks pose significant challenges due to the anonymity, encryption, jurisdictional complexities, technical difficulties, and ethical considerations associated with these networks. Overcoming these challenges requires specialized expertise, international coordination, and careful consideration of privacy and surveillance concerns. Law enforcement agencies need to adapt and develop innovative approaches to effectively combat cybercrime in onion routing networks while safeguarding users' privacy rights and adhering to legal frameworks.

6- COUNTERMEASURES AGAINST CYBERCRIME IN ONION ROUTING NETWORKS

Several countermeasures can be implemented to detect and mitigate cybercrime in onion routing networks. Some of the key countermeasures include:

6.1 **Law Enforcement Cooperation.** International cooperation among law enforcement agencies is essential in detecting and mitigating cybercrime in onion routing networks. This can involve collaboration among different jurisdictions to overcome challenges related to jurisdictional boundaries. Mutual legal assistance treaties (MLATs) and other forms of international cooperation mechanisms can be utilized to facilitate information sharing, evidence gathering, and extradition of cyber criminals. Law enforcement agencies can work together to coordinate efforts in identifying and apprehending cybercriminals involved in illegal activities such as cyber-attacks, fraud, and illicit trade in onion routing networks [24].

**6.2 Advanced Monitoring and Analysis Tools.** Advanced monitoring and analysis tools specifically designed for onion routing networks can help in detecting and mitigating cybercrime activities. These tools can analyze network traffic, identify suspicious patterns or behaviors, and uncover hidden connections among different nodes and users. Advanced analytics, machine learning, and artificial intelligence technologies can also be employed to analyze large volumes of data and identify potential cybercrime activities. These tools can provide insights and alerts to law enforcement agencies, enabling them to take timely action against cybercriminals [25].

**6.3 Enhanced Encryption Techniques**. Enhanced encryption techniques can be implemented to improve the security and privacy of onion routing networks while also enabling effective detection and mitigation of cybercrime. For example, traffic analysis-resistant encryption can be used, where the content of communications is encrypted along with additional metadata to prevent traffic analysis attacks. Enhanced encryption techniques can help protect the privacy of legitimate users while also enabling law enforcement agencies to detect and track illegal activities in onion-routing networks. However, striking a balance between privacy and security is important to ensure that the rights of legitimate users are respected while cybercrime is effectively addressed [63].

**6.4 Collaboration with Technology Providers**. Collaboration with technology providers, such as developers of onion routing networks or online marketplaces, can be beneficial in detecting and mitigating cybercrime. Technology providers can implement security features on their platforms, such as user authentication, transaction monitoring, and content filtering, to prevent illegal activities. Collaboration with technology providers can also involve sharing information about vulnerabilities and patches, as well as providing technical support to law enforcement agencies in investigating cybercrime activities. Regular communication and collaboration between law enforcement agencies and technology providers can contribute to effective cybercrime prevention and mitigation efforts [26].

**6.5 Capacity Building for Law Enforcement Agencies.** Building the capacity of law enforcement agencies in investigating and prosecuting cybercrime in onion routing networks is crucial. This can involve providing specialized training, resources, and tools for law enforcement personnel to effectively detect, investigate, and prosecute cybercrime in onion routing networks. Capacity-building efforts can also include partnerships with academia, the private sector, and civil society organizations to promote knowledge sharing, research, and innovation in cybercrime detection and mitigation. Continuous training and skill development of law enforcement personnel can enhance their capabilities in dealing with the evolving landscape of cybercrime in onion routing networks.

**6.6 Regular Node Verification.** Regular verification of the integrity and security of nodes in the onion routing network can help identify compromised or malicious nodes that may be used for cybercrime activities. Node operators can implement security measures such as strong authentication, regular security audits, and software updates to ensure the trustworthiness of their nodes. Verification mechanisms can also involve monitoring the reputation and behavior of nodes to detect any suspicious activities or anomalies that may indicate cybercrime [27].

**6.7 User Education and Awareness.** Educating and creating awareness among users about the risks, best practices, and security measures in onion routing networks can play a significant role in preventing cybercrime. Users should be educated about the potential threats, such as phishing attacks, malware, and scams, and trained on how to securely use onion routing networks. User awareness campaigns, training programs, and informational resources can be developed to empower users to protect themselves from cybercrime and report any suspicious activities [28].

**6.8 Timely Incident Response.** Having a robust incident response mechanism in place can help in the timely detection and mitigation of cybercrime in onion routing networks. Law enforcement agencies, node operators, and other stakeholders should have established protocols for reporting and responding to cybercrime incidents. This can involve coordinated efforts to investigate and mitigate cybercrime activities, including data breach incidents, ransomware attacks, or illegal content distribution. Timely incident response can prevent further damage and disruption caused by cybercrime activities [29].

**6.9 Legal and Regulatory Frameworks.** Developing and enforcing appropriate legal and regulatory frameworks related to onion routing networks can deter cybercrime activities. Laws and regulations can define the legal boundaries and consequences for cybercrime activities, such as hacking, identity theft, fraud, and illegal content distribution. Legal frameworks can also facilitate cooperation and coordination among law enforcement agencies across jurisdictions, streamline evidence gathering and legal proceedings, and enable effective prosecution of cybercriminals [30].

**6.10 Proactive Threat Intelligence.** Proactive threat intelligence gathering and sharing can aid in the early detection and mitigation of cybercrime in onion routing networks. Stakeholders, such as law enforcement agencies, node operators, and technology providers, can collaborate to collect and share threat intelligence, including known cybercrime tactics, techniques, and procedures (TTPs), indicators of compromise (IOCs), and vulnerabilities. Proactive threat intelligence can enhance the situational awareness and preparedness of stakeholders to detect and respond to emerging cyber threats in onion routing networks [31].

In a nutshell, a multi-faceted approach involving international cooperation, advanced monitoring and analysis tools, enhanced encryption techniques, collaboration with technology providers, capacity building for law enforcement agencies, regular node verification, user education, and awareness, timely incident response, legal and regulatory frameworks, and proactive threat intelligence can be effective in detecting and mitigating cybercrime in onion routing networks. It requires a combination of technical, legal, and collaborative efforts to effectively combat cybercrime in these networks and protect the security and privacy of legitimate users.

## 7- LIMITATIONS OF COUNTERMEASURES

While the countermeasures mentioned above can help mitigate cybercrime threats in onion routing networks, they also have limitations:

**7.1 Privacy concerns.** Onion routing networks are designed to prioritize privacy and anonymity for users, which can make it difficult to monitor and analyze network traffic for detecting cybercrime activities. The encryption and multiple layers of routing paths in onion routing networks can make it challenging to identify malicious activities and track down cybercriminals. Balancing the need for privacy with the requirements of security and cybercrime prevention is a complex task that requires careful consideration of privacy concerns while implementing effective countermeasures [36].

**7.2 Technical challenges.** Implementing countermeasures such as network monitoring, vulnerability management, and traffic analysis in onion routing networks can be technically challenging. The decentralized nature of the network, the use of encryption, and the complexity of routing paths can make it difficult to detect and prevent cybercrime activities. For example, the use of different encryption techniques can make it challenging to decipher the content of network traffic, and the complex routing paths can obscure the origin of malicious activities. Overcoming these technical challenges requires continuous research and development in the field of cybersecurity to develop innovative solutions that can effectively detect and prevent cybercrime in onion routing networks.

**7.3 Legal and jurisdictional issues.** Onion routing networks operate globally, and cybercriminals can operate from different jurisdictions, making it challenging to prosecute and bring them to justice. Legal and jurisdictional issues can hinder effective law enforcement efforts, as different countries may have varying laws and regulations related to cybercrime and privacy. The lack of a consistent legal framework across different jurisdictions can create challenges in investigating and prosecuting cybercriminals operating in onion-routing networks. International cooperation among countries is crucial to combat cybercrime in onion routing networks effectively. Developing effective legal and regulatory frameworks that address cybercrime in onion routing networks while respecting privacy and jurisdictional concerns is a complex task that requires international coordination and cooperation [37].

**7.4 Resource constraints.** Implementing effective countermeasures against cybercrime in onion routing networks requires resources, including technical expertise, funding, and infrastructure. However, not all entities may have the necessary resources to implement robust countermeasures, resulting in varying levels of security and vulnerability among different parts of the network. Small organizations or individuals may lack the technical expertise or funding to implement sophisticated security measures, making them more susceptible to cybercrime attacks. Addressing resource constraints and ensuring that all stakeholders have access to the necessary resources, such as funding for cybersecurity measures and technical expertise, is essential in combating cybercrime in onion routing networks effectively.

**7.5 Human factor vulnerabilities.** Human error and behavior can be a limitation in countering cybercrime in onion routing networks. Users may inadvertently disclose sensitive information or fall victim to social engineering attacks, leading to security breaches. Human factor vulnerabilities can include poor password management, clicking on malicious links, falling for phishing attacks, and other forms of user-related security lapses. Educating and training users about best practices in cybersecurity, promoting awareness about potential threats, and fostering a security-conscious culture among users can help mitigate human factor vulnerabilities. However, human behavior can still be unpredictable and pose challenges in effectively countering cybercrime in onion routing networks [32].

In summary, the limitations of countermeasures against cybercrime in onion routing networks include privacy concerns, technical challenges, legal and jurisdictional issues, resource constraints, and human factor vulnerabilities. Overcoming these limitations requires a holistic approach that balances privacy with security, addresses technical challenges through continuous research and development, promotes international cooperation among countries, and ensures that all stakeholders have access to the necessary resources to implement effective countermeasures.

## 8- FUTURE DIRECTIONS

As cyber threats continue to evolve, the field of onion-routing networks and countermeasures against cybercrime will also need to evolve. Some potential future directions for addressing cybercrime in onion routing networks include:

**8.1 Advanced traffic analysis techniques.** As onion routing networks continue to evolve, there is a need to develop advanced traffic analysis techniques that can effectively analyze network traffic despite encryption and routing paths. This can involve the use of machine learning algorithms, artificial intelligence, and other advanced techniques to analyze patterns of cybercrime activities. For example, developing algorithms that can detect anomalous patterns of behavior, such as sudden spikes in data transfer or unusual communication patterns, can help identify potential cybercrime activities within the encrypted network traffic [35].

**8.2 Enhanced collaboration among stakeholders.** Strengthening collaboration among various stakeholders, including the Tor community, law enforcement agencies, government agencies, and other entities, can lead to more effective strategies for combating cybercrime in onion routing networks. This can involve sharing threat intelligence, coordinating incident response efforts, and developing joint initiatives to improve the overall security of the network. For example, law enforcement agencies can work closely with Tor developers to identify vulnerabilities and develop countermeasures, while the Tor community can provide feedback and insights to help improve the security of the network [38].

**8.3 Improved user education and awareness.** Educating and raising awareness among users about the potential cybercrime threats in onion routing networks can help users take appropriate precautions and avoid falling victim to cybercrime activities. This can include promoting safe browsing practices, providing user-friendly tools and resources for protecting privacy and security, and addressing user concerns and queries. For example, providing regular updates about the latest threats, offering tutorials on how to use Tor securely, and promoting responsible use of the network can empower users to protect themselves and contribute to a safer onion-routing network [39].

**8.4 Strengthened legal and regulatory measures.** Developing and implementing robust legal and regulatory measures that specifically address cybercrime in onion routing networks can provide a legal framework for prosecuting cybercriminals and deterring illegal activities. This may involve international cooperation among countries, harmonization of laws, and addressing jurisdictional challenges. For example, establishing clear laws that define cybercrime in the context of onion routing networks, providing legal mechanisms for law enforcement agencies to investigate and prosecute cybercriminals, and developing international agreements for cooperation in combating cybercrime can help strengthen the legal and regulatory framework for addressing cybercrime in onion routing networks [34].

**8.5 Enhanced resource allocation.** Ensuring that all stakeholders, including Tor developers, network operators, law enforcement agencies, and users, have access to the necessary resources, including technical expertise, funding, and infrastructure, can help strengthen the overall security of onion routing networks. Efforts should be made to address resource gaps and provide support to stakeholders to effectively combat cybercrime. For example, providing funding for research and development of new security technologies, allocating resources for training and capacity building for law enforcement agencies, and promoting public-private partnerships to support cybersecurity initiatives can help enhance resource allocation and improve the security of onion routing networks [33].

In short, addressing cybercrime in onion routing networks requires a multi-faceted approach that involves advanced traffic analysis techniques, enhanced collaboration among stakeholders, improved user education and awareness, strengthened legal and regulatory measures, and enhanced resource allocation. By taking proactive measures and continuously adapting to evolving cyber threats, the field of onion routing networks can better mitigate the risks associated with cybercrime and ensure a safer and more secure online environment.

## 9- DISCUSSION AND RESULTS

The discussion of this research focuses on the types of cybercrime in onion-routing networks, challenges in detecting and mitigating cybercrime, countermeasures against cybercrime, limitations of countermeasures, and future directions for addressing cybercrime in onion-routing networks.

Based on the literature review and analysis of various types of cybercrime in onion routing networks, it is evident that these networks pose significant challenges in detecting and mitigating cybercrime due to the inherent anonymity, encryption, and jurisdictional complexities associated with onion routing. The discussion and results of this research highlight the following key findings:

**9.1 Types of Cybercrime in Onion Routing Networks.** The research identifies various types of cybercrime that can occur in onion routing networks. These include drug trafficking, fraud, hacking, money laundering, weapons trafficking, child exploitation, cyber extortion, cyber espionage, terrorism financing, and counterfeiting. These cybercrimes involve illegal activities such as the sale of illegal drugs, fraudulent schemes, unauthorized access to systems, illicit money transfers, illegal weapons trade, exploitation of minors, extortion through cyber means, espionage activities, funding of terrorism, and production of counterfeit goods. The discussion can further explore the prevalence and impact of each type of cybercrime in onion routing networks, highlighting the challenges associated with detecting and mitigating them.

**9.2 Challenges in Detecting and Mitigating Cybercrime in Onion Routing Networks**. The research identifies several challenges in detecting and mitigating cybercrime in onion routing networks. These challenges include anonymity, encryption, jurisdictional issues, technical challenges, and ethical considerations. The discussion can elaborate on each challenge, discussing the complexities associated with tracing and identifying cybercriminals due to the anonymous nature of onion routing networks, the use of encryption to conceal illegal activities, difficulties in establishing jurisdiction and legal authority, technical challenges in monitoring and analyzing traffic, and ethical considerations related to privacy and data protection. The discussion can also highlight the interplay between these challenges and how they collectively hinder the effective detection and mitigation of cybercrime in onion routing networks.

**9.3 Countermeasures against Cybercrime in Onion Routing Networks.** The research proposes several countermeasures to address cyber-crime in onion routing networks. These countermeasures include law enforcement cooperation, advanced monitoring and analysis tools, enhanced encryption techniques, collaboration with technology providers, capacity building for law enforcement agencies, regular node verification, user education and awareness, legal and regulatory frameworks, and proactive threat intelligence. The discussion can delve into each countermeasure, discussing how they can contribute to mitigating cybercrime in onion routing networks. It can also explore the potential benefits and limitations of each countermeasure, their feasibility, and potential challenges in implementing them effectively.

**9.4 Limitations of Countermeasures.** The research acknowledges several limitations of the proposed countermeasures. These limitations include privacy concerns, technical challenges, legal and jurisdictional issues, resource constraints, and human factor vulnerabilities. The discussion can further explore these limitations, discussing the potential ethical implications of countermeasures that may infringe on privacy rights, technical challenges associated with implementing and maintaining countermeasures, legal and jurisdictional issues related to international cooperation, resource constraints in terms of funding and technical capabilities, and human factor vulnerabilities such as insider threats and social engineering attacks.

**9.5 Future Directions.** The research identifies several future directions for addressing cybercrime in onion routing networks. These include advanced traffic analysis techniques, enhanced collaboration among stakeholders, improved user education and awareness, strengthened legal and regulatory measures, and enhanced resource allocation. The discussion can elaborate on these future directions, discussing how advancements in traffic analysis techniques can aid in detecting and mitigating cybercrime, the importance of collaboration among stakeholders such as law enforcement agencies, technology providers, and policymakers, the need for enhanced user education and awareness to prevent cybercrime, the role of legal and regulatory frameworks in addressing cybercrime in onion routing networks, and the importance of adequate resource allocation to effectively combat cybercrime.

Ultimately, the research concludes by summarizing the discussion and results presented in the study. It highlights the challenges associated with detecting and mitigating cybercrime in onion routing networks, proposes countermeasures to address these challenges, acknowledges the limitations of countermeasures, and suggests future directions for improving cybercrime prevention

## 10- CONCLUSION

In conclusion, the detection and mitigation of cybercrime in onion routing networks present complex challenges due to factors such as anonymity, encryption, jurisdictional complexities, technical complexities, and ethical considerations. However, effective countermeasures can be implemented through various approaches. Law enforcement cooperation at the international level, specialized monitoring and analysis tools, advanced encryption techniques, collaboration with technology providers, and capacity building for law enforcement agencies are key strategies for addressing these challenges.

International cooperation among law enforcement agencies can facilitate information sharing, joint investigations, and extradition of cyber criminals. Specialized monitoring and analysis tools can help in identifying suspicious activities and patterns of behavior within onion routing networks. Advanced encryption techniques can enhance the security of communications and data within these networks. Collaboration with technology providers can involve implementing measures such as enhanced user authentication and data retention policies. Capacity-building efforts can improve the technical skills and knowledge of law enforcement agencies in investigating cybercrime in onion routing networks.

However, it is important to strike a balance between investigating cybercrime and protecting the privacy and security of legitimate users. Ethical considerations related to surveillance and civil liberties must also be addressed. Further research, innovation, and collaboration among stakeholders including law enforcement, technology providers, academia, and civil society are essential in effectively tackling cybercrime threats in onion routing networks. By combining these approaches, we can work towards mitigating the unique challenges posed by cybercrime in onion routing networks and safeguarding the digital world.

## ACKNOWLEDGMENT

The author wishes to express deep gratitude to the various individuals and organizations whose contributions have been instrumental in the completion of this research. Firstly, sincere appreciation goes to cybersecurity researchers and experts for sharing their profound knowledge and insights into onion-routing networks and cybercrime threats. Their expertise has significantly enriched the depth of this study.

Furthermore, the author extends thanks to law enforcement agencies and cybersecurity organizations for providing invaluable data and information related to cybercrime activities within onion routing networks. Their cooperation has been crucial in understanding the complex dynamics of this realm. The Tor Project and its dedicated community members deserve recognition for their continuous efforts in developing and maintaining the Tor network, which forms the core focus of this research. Their commitment has laid the foundation for exploring the intricacies of onion-routing networks.

Additionally, the author acknowledges the invaluable contributions of colleagues and peers who have provided feedback, suggestions, and engaging discussions, thereby enhancing the overall quality of this research. Special appreciation is also extended to funding agencies or organizations whose financial support has facilitated the execution of this research, enabling thorough investigation and analysis.
Lastly, heartfelt gratitude is expressed to all users who prioritize privacy and security while responsibly utilizing onion routing networks. Their conscientious approach contributes to the broader understanding of these networks and their implications.